\documentclass[pre,twocolumn,showpacs,superscriptaddress,preprintnumbers,floatfix]{revtex4-1}

\usepackage{etex}
\usepackage{ifpdf}
\usepackage{hyperref}
\usepackage{dcolumn}
\usepackage{url}
\usepackage{amsmath}
\usepackage{amscd}
\usepackage{amsfonts}
\usepackage{amssymb}
\usepackage{amsthm}
\usepackage{xfrac}
\usepackage{braket}
\usepackage{bm}   
\usepackage{bbm}
\usepackage{verbatim}
\usepackage{stmaryrd}
\usepackage{xcolor}
\usepackage{setspace}
\usepackage{tikz}
\usetikzlibrary{arrows}
\usetikzlibrary{automata}
\usetikzlibrary{positioning}
\usetikzlibrary{plotmarks}
\usepackage{pgfplots}
\pgfplotsset{compat=newest}

\tikzstyle{vaucanson}=[
  node distance=2.5cm,
  bend angle=15,
  auto,
  every loop/.style={},
  every edge/.style={->,draw=black,line width=1.2,>=latex,shorten <=1pt,shorten >=1pt},
  every state/.style={draw=black,line width=2,font=\large},
  loop right/.style={right,out=22,in=-22,loop},
  loop above/.style={above,out=112,in=68,loop},
  loop left/.style={left,out=202,in=158,loop},
  loop below/.style={below,out=292,in=248,loop},
  loop above right/.style={right,out=67,in=23,loop},
  loop above left/.style={left,out=157,in=113,loop},
  loop below left/.style={left,out=247,in=203,loop},
  loop below right/.style={right,out=337,in=293,loop},
]

\theoremstyle{plain}    
\theoremstyle{plain}    
\theoremstyle{plain}    
\theoremstyle{plain}    
\theoremstyle{plain}    
\theoremstyle{plain}    
\theoremstyle{plain}    
\theoremstyle{plain}    
\theoremstyle{plain}    
\theoremstyle{plain}    
\theoremstyle{plain}    
\theoremstyle{plain}    
\theoremstyle{plain}

\newcommand{\eM}     {\mbox{$\epsilon$-machine}}
\newcommand{\eMs}    {\mbox{$\epsilon$-machines}}

\newcommand{\MeasAlphabet}  {\mathcal{A}}
\newcommand{\MeasSymbol}   { {X} }
\newcommand{\meassymbol}   { {x} }

\newcommand{\CausalState}   { \mathcal{S} }

\newcommand{\causalstate}   { \sigma }
\newcommand{\CausalStateSet}    { \boldsymbol{\CausalState} }
\newcommand{\AlternateState}    { \mathcal{R} }

\newcommand{\AlternateStateSet} { \boldsymbol{\AlternateState} }

\newcommand{\Prob}      {\Pr} 

\newcommand{\Cmu}       {C_\mu}
\newcommand{\hmu}       {h_\mu}
\newcommand{\EE}        {\mathbf{E}}

\newcommand{\ProcessAlphabet}   {\MeasAlphabet}

\newcommand{\forward}{+}
\newcommand{\reverse}{-}
\newcommand{\forwardreverse}{\pm} 

\newcommand{\FutureCausalState} { {\CausalState}^{\forward} }

\newcommand{\PastCausalState}   { {\CausalState}^{\reverse} }

\newcommand{\lastindex}[2]{
  \edef\tempa{0}
  \edef\tempb{#2}
  \ifx\tempa\tempb
    \edef\tempc{#1}
  \else
    \edef\tempa{0}
    \edef\tempb{#1}
    \ifx\tempa\tempb
      \edef\tempc{#2}
    \else
      \edef\tempc{#1+#2}
    \fi
  \fi
  \tempc
}

\newcommand{\MOrder}{R}

\newcommand{\CSjoint}[1][,]{
   \edef\tempa{:}
   \edef\tempb{#1}
   \ifx\tempa\tempb
      \ensuremath{\FutureCausalState\!#1\PastCausalState}
   \else
      \ensuremath{\FutureCausalState#1\PastCausalState}
   \fi
}

\newif\ifpm
\edef\tempa{\forwardreverse}
\edef\tempb{\pm}
\ifx\tempa\tempb
   \pmfalse
\else
   \pmtrue
\fi

\colorlet {R_color}    {blue}
\colorlet {k_color}    {black!30!green}

\def\clap#1{\hbox to 0pt{\hss#1\hss}}

\begin{document}

\title{Signatures of Infinity:\\
Nonergodicity and Resource Scaling in\\
Prediction, Complexity, and Learning}

\author{James P. Crutchfield}
\email{chaos@ucdavis.edu}
\affiliation{Complexity Sciences Center and Department of Physics, University of
  California at Davis, One Shields Avenue, Davis, CA 95616}

\author{Sarah Marzen}
\email{smarzen@berkeley.edu}
\affiliation{Department of Physics, University of California at Berkeley,
Berkeley, CA 94720-5800}

\date{\today}
\bibliographystyle{unsrt}

\begin{abstract}
We introduce a simple analysis of the structural complexity of infinite-memory
processes built from random samples of stationary, ergodic finite-memory
component processes. Such processes are familiar from the well known multi-arm
Bandit problem. We contrast our analysis with computation-theoretic and
statistical inference approaches to understanding their complexity. The result
is an alternative view of the relationship between predictability, complexity,
and learning that highlights the distinct ways in which informational and
correlational divergences arise in complex ergodic and nonergodic
processes. We draw out consequences for the resource divergences that
delineate the structural hierarchy of ergodic processes and for
processes that are themselves hierarchical.

\vspace{0.2in}

\end{abstract}

\preprint{Santa Fe Institute Working Paper 15-03-XXX}
\preprint{arxiv.org:1503.XXXX [physics.gen-ph]}

\maketitle


\setstretch{1.1}

\newcommand{\Abet}{\ProcessAlphabet}
\newcommand{\MS}{\MeasSymbol}
\newcommand{\ms}{\meassymbol}
\newcommand{\SSet}{\CausalStateSet}
\newcommand{\St}{\CausalState}
\newcommand{\st}{\causalstate}
\newcommand{\MxSt}{\AlternateState}
\newcommand{\MxSSet}{\AlternateStateSet}
\newcommand{\mxst}{\mu}
\newcommand{\mxstt}[1]{\mu_{#1}}
\newcommand{\StartMS}{\bra{\delta_\pi}}


\section{Introduction}


Truly complex stochastic processes---the \emph{infinitary processes}
\cite{Crut01a} whose mutual information between past and future
diverges---arise in many physical and biological systems
\cite{Crut89e,Bial00a,Trav11b,Debo12a}, such as those in critical states. They
are implicated in many natural phenomena, from the geophysics of earthquakes
\cite{Turc97a} and physiological measurements of neural avalanches
\cite{Begg03a} to semantics in natural language \cite{Debo11a} and cascading
failure in power transmission grids \cite{Dobs07a}. Their apparent infinite
memory makes empirical estimation and modeling particularly challenging. The
difficulty is reflected in the computational complexity of
inference \cite{Kear94a}: the
resources required to predict and model them diverge in sample size, in memory
for storing model parameters, and in memory required for prediction. Resource
scaling, an analog of the venerable technique of finite-size scaling in
statistical mechanics, suggests that for infinitary processes we look for
statistical signatures that track divergences. Since resource divergences are
sensitive to a process's inherent randomness and organization, one hopes that
their scaling forms are uniquely revealing indicators of process complexity and
can guide the selection of appropriate models.

To date, though, there are few tractable constructions with which to explore
possible general relationships between prediction, complexity, and learning for
infinitary processes. One of the few tractable and general constructions is the
class of \emph{Bandit processes} constructed from repeated trials of an
experiment whose properties are, themselves, varying stochastically from trial
to trial \cite{Cove70a,Berr72a}. Even if each individual trial is a realization
generated by a stationary process with finite memory and exponentially decaying
correlations, the resulting process over many trials can be infinitary
\cite{Bial00a,Trav11b,Debo12a}.

Why can the past-future mutual information of Bandit processes diverge?
The answer is remarkably simple: Bandit processes are nonergodic. More to the
point, the divergence is driven by memory in the nonergodic part of their
construction---the mechanism in each trial that selects and then remembers the
operant ergodic component. Here, we use that insight to provide a simple,
alternative derivation of information divergence for this class of infinitary
process: a structural complexity scaling that directly accounts for
nonergodicity.

Information divergence in Bandit processes has been interpreted as reflecting a
universal property of learning: a unique indicator of the number of process
parameters \cite{Bial00a}. The derivation presented here recovers the
connection between the complexity of parameter estimation and divergence in
past-future information.  However, it also identifies other structural
features, such as infinitary ergodic components, that can drive divergences.
Thus, information divergences in Bandit processes reflect particular structural
properties of this class, rather than overarching principles of prediction,
complexity, and learning for infinitary processes. Nonetheless, the issues
raised highlight the need for a more balanced view of truly complex processes
and their challenges. We hope our simplified analysis introduces tools
appropriate to further, detailed scaling analysis of both ergodic and nonergodic
infinitary processes.

Analyzing structural complexity is often conflated with statistical and
computation-theoretic approaches to complex processes. To ameliorate this, the
next section reviews these alternatives. Then we move on to construct Bandit
processes and analyze their structural complexity. We then discuss the results,
draw out contrasts with computation-theoretic and statistical approaches,
highlight the structural hierarchy of ergodic processes, and close with a brief
discussion of hierarchical processes with nested organization.

\section{Prediction, Complexity, and Learning}
\label{sec:PCL}

There is a relationship between, on the one hand, the inherent unpredictability
and memory in a process and, on the other, the difficulty of learning a model
from time series samples and predicting the time series. Alternative framings
lead to different views of this relationship. There are those that attempt to
exactly describe a time series, those that try to express persistent
regularities, and those that consider the consequences for inference. Their
methods are closely related.

The \emph{Kolmogorov-Chaitin complexity} monitors the computational
resources---specifically, length of the minimal program for a given Universal Turing Machine (UTM)---required to reconstruct an individual time series
\cite{Kolm65,Chai66,Mart66a,Kolm83,Brud83,Vita93a}. It is a measure of
randomness: A random time series has no smaller description than itself.
Elaborating on this, \emph{logical depth} \cite{Benn88} and
\emph{sophistication} \cite{Kopp91a} track complementary computational
resources. Logical depth is the number of compute steps the minimal UTM program
requires to generate the time series.  Sophistication is the length of
that part of the UTM program which captures regularities and organization,
effectively discounting the time series' irreducible randomness. All these are
uncomputable, though, even if one is given a generative model.

Fortunately, for a process' typical realizations the Kolmogorov-Chaitin
complexity grows linearly with time series length, with coefficient equal to
\emph{Shannon source entropy rate} $\hmu$ (a measure of a process' unpredictability)
and offset equal to the \emph{statistical complexity} $\Cmu$ (a measure of a process'
memory) \cite[and references therein]{Crut12a}. Given a generative model called
the \emph{\eM}, both the entropy rate and statistical complexity are
computable; if the \eM\ is finite, they are calculable in closed form
\cite{Crut13a}.

We say that $\hmu$, $\Cmu$, and the finite-time excess entropy discussed later are intrinsic measures of a process' structure, randomness, and organization.  By \emph{intrinsic}, we mean that these measures exist independently of the amount of data that we have observed.  The aforementioned algorithmic complexities explicitly depend on the amount of data seen so far, but if the process is ergodic, then algorithmic complexities are also (almost always) intrinsic to a process in the limit of an arbitrarily large amount of data.

Such analyses of intrinsic properties should be contrasted with how statistical
inference approaches complex processes. Statistical learning theory
\cite{Mack03a,Hast11a} analyses and machine learning complexity controls
\cite{Wall87a, Riss89a, Akai74a, Akai77a} are not intrinsic in the sense that
they show how to choose the best in-class model, but the choice of that class
remains subjective. The problem of out-of-class modeling always exists as a
practical necessity, but it is rarely, if ever, tackled directly.  Of course,
in the happy circumstance a correct generative model is in-class, then one has
identified something intrinsic about a process. This, however, begs the
question of discovering the class in the first place. And, practically, such
luck is rarely the case.  Worse, when they do not work well, complexity
controls give no prescription for choosing an alternative class.

Intrinsic complexity characterizations have been most constructively and thoroughly developed for
finite-memory, finite-randomness processes, despite the fact that many
important natural processes are infinitary. The latter include the critical
phenomena \cite{Binn92a} of statistical physics and the routes to chaos in
nonlinear dynamics \cite{Crut89e}, to mention only two. They exhibit
arbitrarily long-range spatiotemporal correlations, infinite memory, and
infinite parameter space dimension. The relationship between prediction,
complexity, and learning is especially interesting when confronted with infinitary
processes, and we re-investigate that relationship for nonergodic Bandit processes.


%

\section{Bandit Process Construction}
\label{sec:2}

The simplest construction of a Bandit process is the following. Consider the
stochastic process generated by a biased coin whose bias $\textbf{P}$ is itself
a random variable. First, a coin bias $p$ is chosen from a user-specified
distribution $\Prob(\textbf{P})$; next, a bi-infinite sequence $\mathbf{x}^1 =
\ms_{-1} \ms_0 \ms_1 \ms_2 \ldots$ is generated from a coin with this
particular bias; then, this is repeated for an arbitrarily large number of such
trials; generating an ensemble $\{\mathbf{x}^1, \mathbf{x}^2, \mathbf{x}^3,
\ldots\}$ of sequences at different biases. The process of interest is this
sequence ensemble. We denote the random variable block between times $a$ and
$b$, but not that at time $b$, as $X_{a:b} = X_a X_{a+1} \ldots X_{b-1}$. We
suppress denoting indices that are infinite. And so, the process of interest is
denoted $X_{:}$. To denote the random variable block conditioned on a random
variable $Z$ taking realization $z$ we use $\MS_{a:b} | Z=z$. So here, the
subprocess $X_{:}|\textbf{P}=p$ is that produced by a coin with bias $p$.

A single one of these bi-infinite sequences comes from an ergodic process that
is memoryless in every sense of the word. In particular, since in each trial
past and future are independent, the conditional past-future mutual information
$I[\MS_{-M:0};\MS_{0:N}|\textbf{P}=p]$ vanishes for any $M$, $N$, and $p$.
However, each of these bi-infinite chains is statistically distinct. The mean
number of heads, say, in one is very different than the mean number of heads in
another. For sufficiently long chains, such differences are almost surely not
the consequence of finite-sample fluctuations.

The overall process $\MS_{:}$ does not distinguish between sequences generated
by different biased coins. So, by making the coin bias a random variable, the
past and future are no longer independent. Both share information about the
underlying coin bias $p$. As we will now show, the shared information or
\emph{excess entropy} $\EE(M,N) = I[\MS_{-M:0};\MS_{0:N}]$ diverges with $M$
and $N$ when $\textbf{P}$ is a continuous random variable.

\section{Information Analysis}
\label{sec:InfoAnal}

To see why, we abstract to a more general case. What follows is an alternative,
direct derivation of results in Ref. \cite[Sec. 4]{Bial00a} that, due to its
simplicity, lends additional transparency to the mechanisms driving the
divergence.

Let $\Theta$ be a random variable with realizations $\theta$ in a (parameter)
space of dimension $K$. $\Theta$ has some as-yet unspecified relationship with
observations $\MS_{:} = \ldots \MS_{-2},\MS_{-1},\MS_0,\MS_1,\ldots$.
We can always perform the following information-theoretic
decomposition of the composite process's excess entropy:
\begin{align}
I[\MS_{-M:0};\MS_{0:N}] & = I[\MS_{-M:0};\MS_{0:N}|\Theta] \nonumber \\
  & \quad + I[\MS_{-M:0};\MS_{0:N};\Theta]
  ~.
\label{eq:1}
\end{align}
The first term quantifies the range of temporal correlations of the observed
process \emph{given} $\Theta$, and the second term quantifies the dependencies
between past and future purely due to $\Theta$. When the fixed-parameter
process $\MS_{:}|\Theta=\theta$ is ergodic and the composite process $\MS_{:}$
is not, then Eq.~(\ref{eq:1}) can be viewed as a decomposition of
$I[\MS_{-M:0};\MS_{0:N}]$ into ergodic and nonergodic contributions, respectively.

The second term $I[\MS_{-M:0};\MS_{0:N};\Theta]$ is a multivariate mutual
information \cite{Jame11a} or \emph{co-information} \cite{Bell03a}.
It is closely related to
parameter estimation, as expected \cite{Bial00a}, since it provides information
about the dimension $K$ of $\Theta$. Standard information-theoretic identities
yield:
\begin{align}
I[\MS_{-M:0};\MS_{0:N};\Theta] & = H[\Theta] + H[\Theta|\MS_{-M:N}] \nonumber \\
  & \quad - H[\Theta|\MS_{-M:0}] - H[\Theta|\MS_{0:N}]
  ~.
\label{eq:2}
\end{align}
The first term $H[\Theta]$ quantifies our intrinsic uncertainty in the bias.
When $\Theta$ is a continuous random variable, $H[\Theta]$ is a differential
entropy. The subsequent terms describe how our uncertainty in $\Theta$
decreases after seeing blocks of lengths $M+N$, $M$, or $N$.

Altogether, Eqs.~(\ref{eq:1}) and (\ref{eq:2}) give:
\begin{align}
I[\MS_{-M:0};\MS_{0:N}]
  & = I[\MS_{-M:0};\MS_{0:N}|\Theta] + H[\Theta] \nonumber \\
  & \quad + H[\Theta|\MS_{-M:N}] - H[\Theta|\MS_{-M:0}] \nonumber \\
  & \quad - H[\Theta|\MS_{0:N}]
  ~.
\label{eq:3}
\end{align}
Thus, assuming one chose a prior with finite entropy $H[\Theta]$, divergences
in $I[\MS_{-M:0};\MS_{0:N}]$ can come from divergences in
$I[\MS_{-M:0};\MS_{0:N}|\Theta]$ or from divergences in $H[\Theta|\MS_{-M:N}] -
H[\Theta|\MS_{-M:0}]- H[\Theta|\MS_{0:N}]$.

Let's take the cases covered in Ref. \cite[Secs. 4.1-4.4]{Bial00a}. There,
$\Theta$ consists of the model parameters, $\theta$ are realizations of
$\Theta$, and $\MS_{:}|\Theta=\theta$ consists of (noisy, potentially
temporally correlated) sequences generated by the model with parameters
$\theta$. For instance, $\Theta$ could be the firing rate of a Poisson neuron
and $\MS_{:}|\Theta=\theta$ could be the time-binned spike trains at firing
rate $\theta$. Or, $\Theta$ could be transition probabilities in a finite
Hidden Markov Model (HMM) and $\MS_{:}|\Theta=\theta$ could be the generated
process given transition probabilities $\theta$. The result, in any case, is a
nonergodic process $\MS_{:}$ constructed from a mixture of ergodic component
processes $\MS_{:}|\Theta=\theta$.

In these examples, the component-process excess entropy
$I[\MS_{-M:0};\MS_{0:N}|\Theta] = \langle
I[\MS_{-M:0};\MS_{0:N}|\Theta=\theta]\rangle_{\theta}$ does not diverge with
$M$ or $N$, since finite HMMs have finite excess entropy, which is bounded by
the internal state entropy \cite{Trav11b,Lohr09a}. In fact, the excess entropy
for many ergodic stochastic processes is finite, even if generated by
infinite-state HMMs. Any divergence in the composite process
$I[\MS_{-M:0};\MS_{0:N}]$ therefore comes from divergences in
$H[\Theta|\MS_{-M:N}]-H[\Theta|\MS_{-M:0}]-H[\Theta|\MS_{0:N}]$. 

Since the composite process includes sequences $\mathbf{\ms}^i$ from trials
with different $\theta$, one's intuition might suggest that
$\Prob(\Theta=\theta|\MS_{-M:0}=\ms_{-M:0})$ is multimodal for most
$\ms_{-M:0}$. However, existing results \cite{walker1969asymptotic,
heyde1979asymptotic, sweeting1992asymptotic, weng2008asymptotic} on the
asymptotic normality of posteriors carry over to this setting, since they
essentially rely on the log-likelihood function $\log
\Prob(\MS_{-M:0}=\ms_{-M:0}|\Theta=\theta)$ being sufficiently well behaved.


For instance, consider the Bandit process construction of Sec.~\ref{sec:2}.
A crude derivation of the asymptotic normality of $\Prob(\Theta=\theta|\MS_{-M:0}=\ms_{-M:0})$ \cite{hartigan1983asymptotic} starts with Bayes Rule:
\begin{align*}
\Prob(\Theta=\theta & | \MS_{-M:0}=\ms_{-M:0}) \\
   & = \frac{\Prob(\MS_{-M:0}=\ms_{-M:0}|\Theta=\theta) 
   \Prob(\Theta=\theta)}{\Prob(\MS_{-M:0}=\ms_{-M:0})}
   ~.
\end{align*}
The denominator $\Prob(\MS_{-M:0}=\ms_{-M:0})$ is quite complicated to calculate, but
this normalization factor does not affect the $\theta$-dependence of
$\Prob(\Theta=\theta|\MS_{-M:0}=\ms_{-M:0})$. More to the point, the prior's
contribution $\Prob(\Theta=\theta)$ is dwarfed by the likelihood:
\begin{align*}
\Prob(|\MS_{-M:0}=\ms_{-M:0} & | \Theta=\theta) \\
  & = \theta^{\sum_{i=0}^{M-1} \ms_i} (1-\theta)^{M-\sum_{i=0}^{M-1} \ms_i}
  ~,
\end{align*}
in the large-$M$ limit.  Let $\theta^*$ be the unique maximum of
$\Prob(\Theta=\theta|\MS_{-M:0}=\ms_{-M:0})$: $\theta^* =
\frac{1}{M}\sum_{i=0}^{M-1} \ms_i + O(1/M)$. Taylor-expanding $\log
\Prob(\Theta=\theta|\MS_{-M:0}=\ms_{-M:0})$ about $\theta^*$ suggests that
$\Prob(\Theta=\theta|\MS_{-M:0}=\ms_{-M:0})$ is approximately normal in the
large-$M$ limit, with variance decaying as $\sim 1/M$. (Any one of the many
sources \cite{walker1969asymptotic, heyde1979asymptotic, sweeting1992asymptotic, weng2008asymptotic}
on asymptotic normality of posteriors provides rigorous and generalized
statements.)

Armed with such asymptotic normality, we now turn our attention to find the
asymptotic form of $H[\Theta|\MS_{-M:0}=\ms_{-M:0}]$,
$H[\Theta|\MS_{0:N}=\ms_{0:N}]$, and $H[\Theta|\MS_{-M:N}=\ms_{-M:N}]$ in the
large-$M$ and -$N$ limits. The differential entropy of a normal distribution
is $\frac{1}{2}\log |\det \Sigma|$, where $\Sigma$ is the covariance matrix;
here, $\det \Sigma \sim 1/M$. This captures the error distribution
for each of the $K$ parameters. So, this and asymptotic normality of the
posterior imply that:
\begin{align*}
H[\Theta|\MS_{-M:0}=\ms_{-M:0}] &\sim -\frac{K}{2}\log M
  ~,
\end{align*}
plus corrections of $O(1)$ in $M$, and thus:
\begin{align*}
H[\Theta|\MS_{-M:0}] &\sim -\frac{K}{2}\log M
  ~,
\end{align*}
where $K$ is the parameter space dimension.

At first blush, the result is counterintuitive. In the limit that $M$ and $N$ tend to
infinity, and we see longer and longer sequences $\ms_{-M:0}$, we become more
certain as to $\Theta$'s value. This increasing certainty should mean
that the conditional entropy $H[\Theta|\MS_{-M:0}=\ms_{-M:0}]$ vanishes.
However, if $\Theta$ is a continuous random variable (such as a Poisson rate),
then $H[\Theta|\MS_{-M:0}=\ms_{-M:0}]$ is a differential entropy.  As our
variance in $\Theta|\MS_{-M:0}=\ms_{-M:0}$ decreases to $0$, the differential
entropy $H[\Theta|\MS_{-M:0}=\ms_{-M:0}]$ diverges to negative infinity. It is
exactly this well known divergence that causes a divergence in
$I[\MS_{-M:0};\MS_{0:N}]$ for the nonergodic processes we are considering.

From these results and Eq. (\ref{eq:3}), one has:
\begin{align*}
I[\MS_{-M:0};\MS_{0:N};\Theta] \sim \frac{K}{2}\log \frac{MN}{M+N}
  ~.
\end{align*}
And, recalling that the ergodic-component information does not diverge,
we immediately recover:
\begin{align}
I[\MS_{-M:0};\MS_{0:N}] \sim \frac{K}{2}\log \frac{MN}{M+N}
  ~.
\end{align}
Lower-order terms in $M$ and $N$ include the expected log-determinant of the
Fisher information matrix for maximum likelihood estimates of $\Theta$
\cite{Lehm98a}.

A similar information-theoretic decomposition can be used to upper-bound the
excess entropy of ergodic processes as well. For instance, App.~\ref{sec:Spin},
uses a similar decomposition to show that the temporal excess entropy of an
Ising spin on a two-dimensional Ising lattice at criticality is finite.




Logarithmic divergences in excess entropy also occur in stationary ergodic
processes, such as exhibited at the onset of chaos through period-doubling
\cite{Crut89e}. And, alternative scalings are known, such as power-law
divergences \cite[Sec. 4.5]{Bial00a}. For natural language texts there is
empirical evidence that the excess entropy diverges. One form is referred to as
Hilberg's Law \cite{Ebel91a,Ebel94a,Debo11a}: $I[\MS_{-N:0};\MS_{0:N}]
\propto \sqrt{N}$.

In contrast with Sec. \ref{sec:InfoAnal}'s rather direct calculation, it is far
less straightforward to analyze these power-law divergences:
\begin{align}
I[\MS_{:0};\MS_{0:N}] \sim N^\gamma
  ~,
\end{align}
with $\gamma \in [0,1)$. While there are results on asymptotics of posteriors
for nonparametric Bayesian inference, many aim to establish asymptotic normality
of the posterior; e.g., as in Refs. \cite{wasserman1998asymptotic,
ghosal2000asymptotic}. As far as we know, no result yet recovers the
aforementioned power-law divergence; likely, since existing asymptotic analyses
avoid the essential singularity for the prior utilized in Ref. \cite[Sec.
4.5]{Bial00a} to obtain power-law divergence.

\section{Discussion}

We investigated one large, but particular class of infinitary processes in
terms of how information measures diverge; recovering, in short order, a
previously reported logarithmic divergence in Bandit-like process past-future
mutual information. Practically, this suggests that one could use the scaling of empirical
estimates of past-future information as a function of sequence length to
estimate a process's parameter space dimension. Mathematically and somewhat
surprisingly, the derivation shows that the reason Bandit-like processes
exhibit information divergences derives from the role nominally finite-sample
effects (asymptotic normality) play in a framework that otherwise assumes
arbitrarily large amounts of data.

\begin{figure}[htp]
\includegraphics[width=.35\columnwidth]{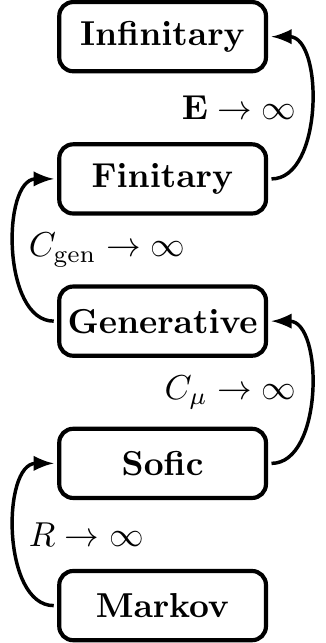}
\caption{Prediction hierarchy for stationary ergodic processes: Each level
  describes a process class with finite informational quantities. A class above
  finitely models the processes in the class below. Classes are separated by
  divergence in the corresponding informational quantity. Moving up the
  hierarchy corresponds to it diverging. Example processes that are finitely
  presented at each level, but infinitely presented at the preceding lower
  level.
  \emph{Sofic}: typical unifilar HMMs, e.g., Even Process \cite{Crut01a};
  \emph{Generative}: typical nonunifilar HMMs \cite{Lohr09a};
  \emph{Finitary}: typical infinite nonunifilar HMMs;
  \emph{Infinitary}: highly atypical infinite HMMs with long-range memory,
  e.g., the ergodic construction in Ref. \cite{Trav11b}.
  }
\label{fig:Hierarchy}
\end{figure}

Section \ref{sec:InfoAnal}'s scaling analysis left open the possibility that
information divergences can be driven by the ergodic components themselves. So,
what is known about information divergences in ergodic processes?  An
information divergence hints at a structural level in the space of ergodic
processes; a space that is itself highly organized. This is seen in the
hierarchy of divergences separating processes into classes of distinct
architecture, depicted in Fig. \ref{fig:Hierarchy}. (See also Table 1, Fig. 18,
and Sec. 5 in Ref. \cite{Crut92c}.) Processes at each level are distinguished
by different scalings for their complexity and in how difficult they are to
learn and predict.


At the lowest level (\emph{Markov}) are processes described by finite \eMs\
with finite history dependence (finite Markov order $\MOrder$); e.g., those
described by existing Maximum Caliber models \cite{Presse13} or by measure
subshifts of finite type \cite{Lind95a}. Though very commonly posited as
models, they inhabit a vanishingly small measure in the space of processes
\cite{Jame10a}. At the next level (\emph{Sofic}) of structure are processes
described by \eMs\ with finite $\Cmu$.  These typically have infinite
Markov order; e.g., the measure-sofic processes.
Above this level
are processes generated by general (that is, nonunifilar) HMMs with uncountable
recurrent causal states and divergent statistical complexity that, nonetheless,
have finite generative complexity, $C_\text{gen} < \infty$ \cite{Lohr09a}.
Processes at the \emph{generative} level not only have infinite Markov order
and storage, but also require a growing amount of memory for accurate
prediction. One consequence is that they are inherently unpredictable by any
observer with finite resources. Note, however, that predictability is
complicated at all levels by \emph{cryptic processes} \cite{Crut08b}---those
with arbitrarily small excess entropy, but large statistical complexity. When
the smallest generative model is infinite but the process still has short-term memory, we arrive at the class of
\emph{finitary} processes ($\EE < \infty$).

Processes with divergent excess entropy---infinitary processes---inhabit the
upper reaches of this hierarchy. Predicting such processes necessarily requires
infinite resources, but accurate prediction can also return infinite
dividends. We agree, here, with Ref. \cite{Bial00a}: the asymptotic rate of
information divergence is a useful proxy for process complexity. Historically,
this view appears to have been anticipated in Shannon's introduction of the
\emph{dimension rate} \cite[App. 7]{Shan48a} of an ensemble of functions:\begin{align*}
\lambda = \lim_{\delta \to 0} \lim_{\epsilon \to 0} \lim_{T \to
\infty}  \frac{N(\epsilon,\delta,T)}{T \log \epsilon}
  ~,
\end{align*}
where $N(\epsilon,\delta,T)$ is the smallest number of elements that can be
chosen such that all elements of the ensemble, apart from a set of measure
$\delta$, are within the distance $\epsilon$ of at least one of those chosen. 

However, it is as important to know which process mechanism drives the
divergence as it is to know the divergence rate. Infinitary Bandit processes
store memory entirely in their nonergodic component. Our analysis identified
the divergence in this memory with the well known divergence in the
differential entropy of highly peaked distributions of vanishing width.
Generalizing Bandit processes to have \emph{structured} ergodic components, we
now see that even finite \eMs\ trivially generate infinitary processes when
their transition probabilities are continuous random variables.

Thus, in this case, we also agree that information divergence is a ``necessary
but not sufficient'' criteria for process complexity \cite{Debo12a}. (Appendix
\ref{sec:Spin}, however, looks at critical phenomena in spin systems to call
out a caveat.) This leaves open a broad challenge to understand the sufficient
mechanisms for information divergences. For example, we have yet to develop
similar informational and computation-theoretic analyses for the infinitary
ergodic processes in Refs. \cite{Trav11b,Debo12a}.

Looking forward, the simplicity of our structural complexity analysis opens up
the possibility to better frame information in hierarchical processes
\cite[Sec. 5]{Crut92c}, such as the structural hierarchy in biology \cite[Fig.
6]{Shen06a}, epochal evolution \cite{Nimw96a}, and knowledge hierarchies in
social systems such as semantics in human language \cite{Chom56}. These are
processes in which multiple levels of mechanism are manifest and operate
simultaneously and in which each level is separated from those below via
phase transitions that lead to various signatures of informational and
structural divergence.


\vspace{-0.1in}
\acknowledgments
\vspace{-0.1in}

We thank N. Ay, W. Bialek, R. D'Souza, C. Ellison, C. Hillar, and I. Nemenman for useful conversations.
The authors thank the Santa Fe Institute for its hospitality during visits.
JPC is an SFI External Faculty member.
This material is based upon work supported by, or in part by, the
U. S. Army Research Laboratory and the U. S. Army Research Office under
contracts W911NF-12-1-0288, W911NF-13-1-0390, and W911NF-13-1-0340.
SM was funded by a National Science Foundation Graduate Student Research Fellowship and the U.C. Berkeley Chancellor's Fellowship.


\vspace{-0.1in}
\appendix
\vspace{-0.1in}

\section{Truly Complex Spin Systems?}
\label{sec:Spin}
\vspace{-0.1in}

Reference \cite{Debo12a} pointed out that many infinitary processes do not
satisfy intuitive definitions for complexity. It suggested that divergence in
$\EE$ is a ``necessary but not sufficient condition'' for a process being truly
complex. While intuitively compelling, perhaps divergent $\EE$ is not even a necessary condition. Let's explain.

Spin systems at criticality are one of the most familiar examples of truly
complex processes: global correlations emerge from purely local interactions
\cite{Binn92a}. Evidence of this complexity appears even if we are only allowed
to observe a single spin's interaction with another on the lattice. At the
critical temperature, the interaction has a power-law autocorrelation function;
at all other temperatures, the spin's autocorrelation function is
asymptotically exponential. The \emph{spatial} excess entropy of these
configurations appears to diverge at criticality \cite{Lau12a}, too. However,
does the \emph{temporal} excess entropy $\EE(M,N)$---roughly, the interaction a
single spin with itself at later times---also diverge at criticality?

Surprisingly, the excess entropy of the dynamics of a single spin on an Ising
lattice is finite, even at the critical temperature, \emph{unless} there are
nonlocal spatial interactions between lattice spins. Consider evolving the
lattice configurations via Glauber dynamics for concreteness \cite{Binn92a}.
That is, spin $j$'s next state $\sigma^j_{t+1}$ is determined stochastically by
its previous state $\sigma^j_t$ and its effective magnetic field $h^j_t =
\sum_{i} J_{ij} \sigma^i_t$. In other words, $h^j_t$ and $\sigma^j_t$ causally
shield the past $\overleftarrow{\sigma}^j_t$ from the future
$\overrightarrow{\sigma}^j_t$, implying that:
\begin{align*}
I[\sigma^j_{t-M:t}; \sigma^j_{t+1:t+N}|h^j_t]
  & = I[\sigma^j_t;\sigma^j_{t+1} | h^j_t] \\
  & \leq H[\sigma^j_t]
  ~.
\end{align*}
Given a finite set of spin values and local interactions, $h^j_t$ can only take
a finite number of values. Thus, $H[h^j_t] < \infty$, and so:
\begin{align*}
\big\vert I[\sigma^j_{t-M:t}; \sigma^j_{t+1:t+N};h^j_t] \big\vert
  & \leq H[h^j_t] \\
  & < \infty
  ~,
\end{align*}
as well.

A more familiar example makes this concrete. For the standard two-dimensional
Ising lattice $J_{ij} = J$, if $i$ and $j$ are nearest neighbors, and $J_{ij} =
0$, otherwise. There, $h^j_t$ can only take $5$ possible
values---$h^j \in \{0,~J,~2J,~3J$, and $4J\}$---giving:
\begin{align*}
\big\vert I[\sigma^j_{t-M:t}; \sigma^j_{t+1:t+N};h^j_t] \big\vert
  & \leq H[h^j_t] \\
  & \leq \log_2 5 ~\text{bits}
  ~.
\end{align*}

The information-theoretic decomposition in Eq.~(\ref{eq:1}) applies in this
particular situation. Here, observed variables $\MS_t$ are spins $\sigma_t$, and
the parameters $\Theta$ are replaced by $h^j$. The bounds above then directly
imply that $\EE(M,N) < \infty$ for all $M$ and $N$. In fact, for the standard
two-dimensional Ising lattice, we find that $\EE(-\infty,\infty) \leq 1 +\log_2
5 = 3.4$ bits. We expect excess entropy to diverge only when $h^j$ is a
continuous random variable. This can happen when $J_{ij}$ is nonzero for an
infinite number of $i$'s. However, this necessitates global, not local,
spin-spin couplings.

On the one hand, this analysis does not negate $\EE$'s utility as a generalized
order parameter \cite{Feld98a}. It is still likely maximized at the critical
point, even if its temporal version does not diverge. On the other, our
analysis shows that phenomena---here, spin lattices with purely local
couplings---do not necessarily have divergent $\EE$ even when many would
consider their dynamics to be truly complex when the system is critical.

At first glance, the analysis contradicts the experiments in Fig. $1$ of Ref.
\cite{Bial00a} for the Ising lattice with only local interactions. A more
careful look reveals that there is no contradiction at all. There, coupling
strengths were randomly changed every $400,000$ iterations, so the resultant time
series looked like a concatenation of samples from a Bandit
process. The analysis in Sec.~\ref{sec:InfoAnal} then predicts the
observed logarithmic scaling in Fig. $1$ there for $N \lesssim 25$. However, it also
implies that $\EE(-\infty,N)$ will stop increasing logarithmically at or before
$N = 400,000$.

\bibliography{chaos}

\end{document}